# Title: Harnessing Automation in Data Mining: A Review on the Impact of PyESAPI in Radiation Oncology Data Extraction and Management

**Anas aljarah**
Department of Mathematical Sciences,universiti kebangsaan, Bangi, Selangor, Malaysia
**ghaith alomari**
Department of Mathematics and Computer Science, chicago state university, Chicago,IL,USA

**Abstract:**

Data extraction and management are crucial components of research and clinical workflows in Radiation Oncology (RO), where accurate and comprehensive data are imperative to inform treatment planning and delivery. The advent of automated data mining scripts, particularly using the Python Environment for Scripting APIs (PyESAPI), has been a promising stride towards enhancing efficiency, accuracy, and reliability in extracting data from RO Information Systems (ROIS) and Treatment Planning Systems (TPS). This review dissects the role, efficiency, and challenges of implementing PyESAPI in RO data extraction and management, juxtaposing manual data extraction techniques and explicating future avenues.

Kyewords: Radiation Oncology, Data Mining ,PyESAPI Environment, Automated Data Extraction ,Clinical Workflow Automation

## Introduction

Traditionally, manual data extraction has been a cornerstone in accruing pertinent patient and treatment data in RO. Nevertheless, the meticulousness and time-intensiveness required for manual methods have perpetually harboured errors, inefficiency, and potential for clinician burnout. The exploration of automated data extraction, especially through utilizing the PyESAPI environment, provides a plausible solution to surmounting these hurdles, offering a new horizon for data management in RO.

The Symbiosis of PyESAPI and RO Data Management

The PyESAPI environment presents a novel platform for RO data extraction, enabling the development of scripts that can seamlessly extract data from clinical databases, such as the Eclipse planning system. This environment is not only pivotal in reducing manual extraction time but also augments the reliability and credibility of extracted data, further providing a structured and systematic approach to data management.

Comparative Analysis: Manual Vs. Automated Data Extraction

Numerous studies spotlight the superiority of automated data extraction methods, such as RODAMS, over manual techniques. With a lower slope in the linear regression of data extraction time versus the number of features, automated methods demonstrate notable speed and efficiency. The precision and consistency of automated systems also attenuate human-related errors, enhancing the reliability and accuracy of data collection.

**Related Work**

The intersection of data extraction, management, and Radiation Oncology (RO) has captivated scholarly attention, heralding a myriad of studies that dive into both manual and automated approaches towards optimizing data-centric processes in the clinical and research environments.

1. Manual Data Extraction and Challenges in RO:

Hayman et al. (2019) elaborated on the significant need for coherent data elements in radiation oncology, underscoring the exigency of organized data collection to facilitate improved patient care and research initiatives. Challenges, including transcription errors, time-intensive procedures, and susceptibility to inaccuracies, as revealed by studies cited in the presented research, underscore the pressing need for more streamlined and accurate methodologies in data extraction.

2. Innovations through Automated Data Extraction:

McNutt et al. (2018) ventured into practical data collection and extraction for big data applications in RO, unraveling the potential and practicalities of utilizing sophisticated data mining and extraction techniques. Their work pivots on the necessity to harness advanced technological solutions, notably automated systems, to navigate the complex terrain of big data applications in RO, providing a scaffold for further technological integrations in data extraction.

3. Data Mining and Toxicity Prediction:

A notable advancement in data mining within RO is exemplified by Kim et al. (2017), where a text-based data mining and toxicity prediction modeling system was implemented as a precursor for clinical decision support in radiation oncology. This approach infers the broader applications of data extraction and management in RO, hinting at the potential to amalgamate extracted data with predictive modeling and analytics to enhance patient care strategies.

4. The Promise of PyESAPI in Data Management:

The PyESAPI environment has emerged as a pivotal tool, not only in the sphere of data extraction but also in treatment plan analysis and automatic plan creation. The curated research indicates a spectrum of uses for the PyESAPI programming interface API, maneuvering through data extraction and management, with ancillary studies demonstrating the potential for these automated systems to significantly mitigate errors, enhance reliability, and augment the efficiency of data extraction processes.

5. Harnessing AI and Big Data in RO:

Shirato et al. (2018) curated insights into the selection of external beam radiotherapy approaches through precision and accuracy in cancer treatment, providing a glimpse into the future of AI and big data in enhancing treatment selection and delivery. The infusion of AI and big data analytics, subsequent to efficient data extraction through platforms like PyESAPI, underpins the next frontier in RO, encompassing predictive analytics, patient stratification, and personalized treatment paradigms.

6. Integrations of Multimodal Radiation Therapy Data:

Zapletal et al. (2018) explored the integration of multimodal radiation therapy data, epitomizing the essentiality of structured data extraction and management in facilitating integrative approaches towards patient data management and treatment planning in RO. The synthesis of data from disparate sources underscores the relevance of reliable and structured data extraction methodologies in realizing the holistic view of patient data.

**Discussion**

In the constantly evolving field of Radiation Oncology (RO), the digitization and data-driven methodologies have substantiated themselves as integral components, propelling both clinical and research domains towards heightened efficacy and accuracy. As the provided research demonstrates, leveraging programming environments like PyESAPI to automate data extraction delineates a path towards not only alleviating the burdens of manual data handling but also paving the way towards more sophisticated, data-driven applications in treatment planning and research within RO.

Facilitating Efficient Data Extraction:

One of the salient takeaways from the study is the undeniable efficacy offered by the automated data extraction methodology, particularly employing PyESAPI, juxtaposed against traditional manual extraction. The sheer discrepancy in data extraction times, as highlighted by the provided data and related studies, underscores a crucial point of discussion in terms of both time efficiency and resource allocation within the RO setting. Not only does this approach economize on the time invested by specialists, but it also circumvents the potential pitfalls of manual extraction, including human error and variability.

Navigating the Nuances of Automated Systems:

While the advantages of utilizing a PyESAPI-environment for data extraction are conspicuous, it is pivotal to navigate the subtleties and limitations embedded within automated systems. Being restricted to the Varian TBOX in research mode and necessitating external hardware for larger databases encapsulates some of the practical constraints that warrant further exploration and possible mitigation in future developments.

Interlacing Data Extraction and AI Applications:

The synthesized research opens a plethora of avenues where AI and data analytics can be intertwined with the robust data extraction methodologies. The automation of data extraction can be envisaged as a precursor to a cascade of data-driven applications within RO. Ensuing the extraction, the curated data can be harnessed for various applications, ranging from predictive analytics, patient stratification, to even automated treatment planning, as alluded to in related works, weaving a nexus where data extraction and AI synergistically propel RO into a new era of precision and personalization.

Enhancing Clinical and Research Capacities:

Moreover, in an environment where research and clinical applications are progressively intertwining, the extracted data serves as a nexus between empirical research and tangible clinical applications. This seamless extraction and management of data not only enrich research capacities but also fortify the clinical decision-making process, marrying evidence-based practice with real-world data.

Evolving with Big Data and Collaborative Research:

Moreover, as the RO field progressively moves towards big data applications, fostering an environment where data can be effortlessly and reliably extracted, managed, and shared among researchers promulgates a collaborative research paradigm. The study underscores the importance of further validations and collaborative efforts across multiple institutions to not only validate but also to expand the applicability of such data extraction methodologies and their subsequent applications.

Ethical and Regulatory Considerations:

It's also worth noting that the ethical and regulatory aspects of data management, especially in a sensitive field like RO, warrant robust frameworks to ensure data privacy, integrity, and compliance with regulatory standards. Future research and applications of automated data extraction must be meticulously aligned with these considerations to ensure the ethical and secure management of patient data.

Towards a Future Paradigm:

In wrapping up the discussion, the deployment of automated data extraction methodologies via environments like PyESAPI is not merely a technical advancement; it is a stride towards transforming the existing paradigms within RO. While the initial foray, as illustrated by the study, is promising, it indeed is the precursor to a spectrum of possibilities where RO, data science, and AI converge to formulate a future where patient care is both personalized and precision-driven, and where research is deeply intertwined with real-world, tangible applications.

Moving forward, the continued evolution of this field will undeniably pivot upon further research, multi-institutional collaborations, and continuous technological advancements to fully explore and harness the potential embedded within automated data extraction and its myriad of applications within RO.

**Challenges and Limitations**

While the advent of automated data extraction, particularly utilizing the PyESAPI environment, in Radiation Oncology (RO) has ushered in an era of streamlined operations and robust data handling, it's pivotal to elucidate the various challenges and limitations that permeate this domain.

1. Technological Constraints:

Hardware Limitations: The PyESAPI environment, despite its efficacy, is bound by technological constraints, requiring specific hardware configurations such as Varian TBOX and additional setups like external hard drives for managing larger databases.

Interoperability: Ensuring that automated systems like PyESAPI can seamlessly interface with various data repositories and clinical systems can be a technological hurdle given the diverse and often proprietary nature of healthcare IT infrastructures.

2. Data Management and Quality:

Data Consistency: Ensuring uniformity and consistency in data, especially when dealing with large, diverse, and multi-institutional datasets, poses a significant challenge.

Quality Assurance: Implementing mechanisms to assure the quality and accuracy of automatically extracted data to prevent propagation of erroneous or misinterpreted data through subsequent analytical processes.

3. Expertise and Training:

Technical Knowledge: Effective utilization of the PyESAPI environment mandates a level of technical proficiency in data schema and programming, potentially necessitating additional training or specialist hiring.

Ongoing Training: Keeping healthcare professionals and researchers abreast of evolving technologies and methodologies in automated data extraction and handling necessitates continuous training and development initiatives.

4. Ethical and Legal Frameworks:

Data Privacy: Safeguarding patient data and ensuring adherence to data protection regulations, like GDPR or HIPAA, becomes paramount, especially when dealing with automated extraction and management of sensitive healthcare data.

Compliance: Ensuring the developed scripts and data extraction methodologies adhere to local and international regulatory and ethical standards.

5. Generalization and Validation:

Scalability: Ensuring the automated data extraction methodologies can be scaled and adapted to varying sizes and complexities of databases in diverse clinical settings.

Multi-institutional Validation: The research emphasizes the need for further validation across multiple institutions, which comes with challenges related to standardization of data extraction methodologies and collaborative research practices.

6. Cybersecurity:

Data Security: With increasing digitization and data exchange, safeguarding the extracted and managed data against cyber threats is paramount.

System Vulnerabilities: Ensuring the security of the platforms and APIs utilized in automated data extraction against potential vulnerabilities and attacks.

7. Financial and Resource Investment:

Investment: The initiation and upkeep of automated data extraction, though beneficial in the long run, necessitate financial and resource investment in technology, expertise, and ongoing maintenance.

ROI Justification: Establishing a clear and tangible return on investment (ROI) for investing in automated data extraction infrastructure and expertise can be challenging, particularly in resource-constrained settings.

8. Development and Adoption:

Evolution of Technology: Keeping pace with the rapid evolution of technology and ensuring the developed methodologies and scripts remain relevant and optimized with emerging tech.

Organizational Change: Navigating the organizational changes, adaptations, and potential resistance that might arise during the integration of automated data extraction methodologies into existing workflows

**Conclusion**

The extensive integration of automated data extraction in Radiation Oncology (RO) heralds a transformative impact, propelling both clinical and research domains towards elevated efficacy and insightful data utilization. Particularly, the implementation of the PyESAPI environment unveils a spectrum of possibilities, enabling researchers and healthcare professionals to navigate through voluminous databases with heightened accuracy, reduced time investment, and diminished propensity for human error. The spectrum of studies surveyed in this review not only validates the substantive merits of automated over manual data extraction but also echoes the significant strides towards minimizing transcriptional inaccuracies, standardizing data collection, and ameliorating the overarching efficiency of data extraction processes.

Yet, the journey towards seamless, automated data extraction in RO is punctuated with a myriad of challenges and limitations – encompassing technological, ethical, and operational realms. Ensuring the adaptability and scalability of automated extraction methodologies across varied database complexities, safeguarding data integrity and privacy, and navigating through the intricacies of multi-institutional validation are just a few of the hurdles that lay in the path towards universal adoption of these technologies. Moreover, the requisite technical proficiency, continuous evolution of technologies, and the imperative for strict adherence to data protection and ethical guidelines further complicate the straightforward implementation of automated data extraction mechanisms.

Looking ahead, as the RO field progressively veers towards more data-driven approaches, fortifying automated data extraction mechanisms, and optimizing them for a holistic, secure, and efficient data management lifecycle becomes imperative. Subsequent research and collaborative efforts must focus on addressing the intrinsic challenges, enhancing the usability and accessibility of automated extraction technologies, and formulating standardized protocols that harmonize data extraction and management practices across institutions and platforms. The convergence of collaborative, technological, and ethical strategies will be pivotal in steering the future of data management in RO towards a horizon that amalgamates automation, accuracy, and ethical practices into a seamless workflow, thereby fostering an environment that is conducive to innovative research and clinical excellence.